\documentclass[12pt]{iopart}
\usepackage{setstack}
\usepackage{graphicx}
\usepackage{iopams}

\begin{document}

\title{Entropy production along nonequilibrium quantum jump trajectories}

\author{Jordan M.~Horowitz$^1$ and Juan M.~R.~Parrondo$^2$}

\address{$^1$ Department of Physics, University of Massachusetts at Boston, Boston, MA 02125, USA}
\address{$^2$ Departamento de F\'isica At\'omica, Molecular y Nuclear and GISC, Universidad Complutense de Madrid, 28040 Madrid, Spain}

\ead{jordan.horowitz@umb.edu}

\date{\today}

\begin{abstract}
For classical nonequilibrium systems, the separation of the total entropy production into the adiabatic and nonadiabatic contributions is useful for understanding irreversibility in nonequilibrium thermodynamics.
In this article, we formulate quantum analogs for driven open quantum systems describable by quantum jump trajectories by applying a quantum stochastic thermodynamics.
Our main conclusions are based on a quantum formulation of the local detailed balance condition.
\end{abstract}


\maketitle

\section{Introduction}

  Entropy production has been customarily split into two pieces,
a part due to the system $\Delta S$ and a contribution from the environment $\Delta S_{\rm env}$: $\Delta S_{\rm tot}=\Delta S+\Delta S_{\rm env}$~\cite{deGroot}.
However, the development and subsequent analysis of the fluctuation theorems testify to the value of a distinct decomposition into adiabatic $\Delta S_{\rm a}$ and nonadiabatic $\Delta S_{\rm na}$ contributions~\cite{Esposito2010, Hatano1999, Hatano2001,Speck2005b}
\begin{equation}\label{eq:ent2}
\Delta S_{\rm tot}=\Delta S_{\rm a}+\Delta S_{\rm na}.
\end{equation}
The nomenclature emphasizes that for an adiabatic (slow) process -- during which the system remains in its instantaneous stationary state -- $\Delta S_{\rm na}=0$, and all of the entropy production is due to the adiabatic component, $\Delta S_{\rm tot}=\Delta S_{\rm a}$.
Far from being a simple recasting, this decomposition provides a refined understanding of irreversibility in nonequilibrium processes~\cite{Seifert2008}.
Both $\Delta S_{\rm a}$ and $\Delta S_{\rm na}$ are individually always positive (on average), unlike $\Delta S$ and $\Delta S_{\rm env}$.
Equation~(\ref{eq:ent2}) is particularly useful when applied to stationary states that support dissipative currents.
In this case, $\Delta S_{\rm a}$ quantifies the entropy production needed to maintain these currents.
Whereas, for transitions between stationary states, $\Delta S_{\rm na}$ is a measure of irreversibility that remains finite in the limit of slow switching; $\Delta S_{\rm tot}$ and $\Delta S_{\rm a}$ are essentially useless, as they diverge due to the continuous dissipation present in nonequilibrium stationary states.

Remarkably, each of these entropy productions satisfies a detailed fluctuation theorem on the level of individual, fluctuating, microscopic trajectories.
Namely, the change in trajectory-dependent  total entropy $\Delta s_{\rm tot}$, nonadiabatic entropy $\Delta s_{\rm na}$, and adiabatic entropy $\Delta s_{\rm a}$, can be deduced  by comparing the probability ${\mathcal P}$ of observing a microscopic trajectory $\gamma$ to the probability of its reverse $\tilde\gamma$ occurring in  a distinct thermodynamic process:
\begin{equation}\label{eq:dftTot}
\Delta s_{\rm tot}[\gamma]=\ln\frac{{\mathcal P}[\gamma]}{\tilde{\mathcal P}[\tilde\gamma]},
\end{equation}
\begin{equation}\label{eq:dftNA}
\Delta s_{\rm na}[\gamma]=\ln \frac{{\mathcal P}[\gamma]}{\tilde{\mathcal P}^+[\tilde\gamma]},
\qquad
\Delta s_{\rm a}[\gamma]=\ln \frac{{\mathcal P}[\gamma]}{{\mathcal P}^+[\gamma]},
\end{equation}
where the probability densities $\tilde{\mathcal P}$ and ${\mathcal P}^+$ correspond to different notions of time reversal whose definitions will be elaborated later.
The advantage of representing these entropy productions as ratios of trajectory probabilities is that it immediately implies that they each satisfy an integral fluctuation theorem, and each are positive on average~\cite{Esposito2010}: $\Delta S_{\rm tot}=\langle \Delta s_{\rm tot}\rangle \ge 0$, $\Delta S_{\rm na}=\langle \Delta s_{\rm na}\rangle \ge 0$, and $\Delta S_{\rm a}=\langle \Delta s_{\rm a}\rangle \ge 0$, where the angle brackets denote an average over all trajectories.

Since the discovery of the fluctuation theorems, extending them, and subsequently \eref{eq:ent2}, to a quantum setting has been an active pursuit~\cite{Esposito2009,Campisi2010b}.
Still, a decomposition akin to \eref{eq:ent2} for quantum thermodynamics is lacking.
In this article, we develop such a decomposition.
Our approach is to take \eref{eq:dftTot} and \eref{eq:dftNA} as defining equations for the various entropy productions.
However, adapting them to a quantum setting requires a consistent interpretation of a trajectory for an open quantum system.
The trajectories we analyze here are the individual realizations of a quantum Markov jump process of  a finite-dimensional open quantum system.
In particular, we model the evolution of an open quantum system by employing the repeated interaction method originally proposed in quantum optics~\cite{Kist1999,Spehner2002,Attal2006,Attal2007}.
Within this formalism the effect of an infinite environment is modeled as a rapid sequence of interactions with a series of copies of one small quantum system.
A prototypical example from quantum optics is a series of two-level atoms  that by rapidly passing through a superconducting cavity can act as a thermal reservoir for the confined electromagnetic radiation~\cite{Santos2011}.
After each copy interacts, we measure it, allowing us to indirectly monitor the evolution of our system of interest.
Roughly speaking, this is the least intrusive method of observation, since the measurement, being made after the interaction, cannot affect the system due to causality.
Moreover, by monitoring the environment we have direct access to the energy that flows into the surroundings as heat.
This allows us the capability to not only analyze entropy production, but also the energy balance between work, heat, and internal energy in line with the quantum stochastic thermodynamics introduced by one of us in \cite{Horowitz2012}.

Derivations of fluctuation theorems based on monitoring the environment, as we do here, were realized previously by Derezi\'nski, De Roeck and Maes for heat currents in steady state~\cite{DeRoeck2006,Derezinski2008}; as well as by Crooks for the work done on a weakly-coupled Markovian open quantum system isothermally driven away from equilibrium~\cite{Crooks2008b}; and most recently by Hekking and Pekola for the work dissipated in an isothermally driven two-level system~\cite{Hekking2013}.
The novelty of the present endeavor is to  explicitly address driven systems in conjunction with dissipative currents.
Moreover, by carefully modeling the system-environment interaction in the Markovian limit, we are led to a robust formula for environment entropy production~[\eref{eq:envJump} below] that applies not only to equilibrium environments, but also to more general surroundings -- such as the coherent thermal reservoirs that can induce lasing without inversion~\cite{Scully2003}.

We begin our analysis in \sref{sec:setup} by introducing our system of interest and defining notation.
Then in sections \ref{sec:diag} and \ref{sec:continuous} conditions are introduced and analyzed for the total entropy production to satisfy a detailed fluctuation theorem within the repeated interaction framework in general and specifically for quantum Markov jump processes. 
Nonadiabatic and adiabatic entropy production are discussed for quantum jump processes in \sref{sec:nonadiabatic}, before concluding in \sref{sec:conclusion}.

\section{Setup}\label{sec:setup}

\subsection{Repeated interaction framework}

We have in mind a finite-dimensional quantum system whose density matrix at time $t$ we denote $\rho_t$.
Its time-reversal invariant Hamiltonian $H(\lambda)$ is parameterized by a collection of externally controlled parameters $\lambda$, which we vary with time to do work on the system.
The surroundings are modeled by coupling the system to a rapid sequence of $N$ copies of a finite-dimensional quantum system with Hilbert space $\mathcal{H}_E$ and time-reversal invariant Hamiltonian $H_E=\sum_{l} \varepsilon_{l} \Pi_l$, where $\varepsilon_l\neq \varepsilon_k$ for $l\neq k$ and each $\Pi_l$ projects onto the associated energy eigenspace ${\mathcal H}_l$. 
To be concise, we will call each such copy an \emph{environment}.
Prior to interacting with the system, each environment is with probability $p_l$ in a known density matrix $r_l$ whose support is confined to ${\mathcal H}_l$.
For example, in an equilibrium environment at inverse temperature $\beta$ each state occurs with Boltzmann weight, so each $r_l=\Pi_l/d_l$ is diagonal in the energy basis with $d_l$ the dimension of the $l$ energy eigenspace, and $p_l\propto d_l e^{-\beta \varepsilon_l}$.
Each copy  interacts with the system for a time $\tau$ through the interaction
\begin{equation}\label{eq:V}
V(\lambda)=h\sum_{k\neq l}S_{kl}(\lambda)\otimes R_{kl},
\end{equation}
which is mediated by environment operators $R_{kl}$ ($k\neq l)$ that map vectors from one energy eigenspace ${\mathcal H}_{l}$ to another ${\mathcal H}_k$ and that we require to obey the symmetry $R_{kl}=R_{lk}^\dag$.
Here, $h$ is the coupling strength, and $S_{kl}(\lambda)$ are parameter-dependent system operators  that verify $S_{kl}=S^\dag_{lk}$, since $V$ must be hermitian.
After the interaction, we make an ideal projective measurement to determine whether the environment remained in ${\mathcal H}_l$ or transitioned to another eigenspace ${\mathcal H}_k$.
By recording transitions between energy subspaces, we track the total amount of energy that flows into the surroundings as heat.
This is a minimal requirement to be able to formulate a first-law-like energy balance equation for a quantum stochastic thermodynamics~\cite{Horowitz2012}.

The time-reversal invariance of $H(\lambda)$, $H_E$, and $V$ assumed above is imposed by requiring that they each commute with the time-reversal operator $\Theta$, which is an an anti-unitary operator that satisfies $\Theta^2=I$.
This further implies that the energy eigenstates of $H(\lambda)$ and $H_E$ do not change under the action of $\Theta$ and $[\Theta,R_{kl}]=0$.

\subsection{Quantum thermodynamic process}

Within the framework of stochastic thermodynamics, the entropy production is connected to the ratio of the probabilities to observe different trajectories along distinct thermodynamic processes, as can be seen in \eref{eq:dftTot} and \eref{eq:dftNA}.
Therefore, in order to discern the quantum entropy production, we first review the class of quantum thermodynamic processes  appropriate for the repeated interaction setup, allowing for driving by an external agent and possibly nonequilibrium environments.
This class of processes is inspired by the cavity quantum electrodynamics experiments proposed in~\cite{Kist1999,Santos2011,Ueda1992}.

The \emph{forward} process begins by preparing the initial ensemble diagonal in the eigenbasis of an observable $A$ with non-degenerate eigenvalues $a$, $\rho_A\equiv\sum_a P_a |a\rangle\langle a|$.
After randomly selecting a system from this ensemble, we measure $A$ to confirm the initial state $|a\rangle$.
Then from $t=0$ to $T=N\tau$, we drive the system with protocol $\Lambda=\{\lambda_t\}_{t=0}^T$, while it interacts with the sequence of environments.
During the $k$-th step ($k=1,\dots, N$) from $t=(k-1)\tau$ to $k\tau$, an environment initially in the ${\mathcal H}_{i_k}$ subspace with density matrix $r_{i_k}$ will end the interval in ${\mathcal H}_ {j_k}$ (where it may be $j_k=i_k$).
Meanwhile, the system density matrix $\rho_{k-1}\equiv\rho_{(k-1)\tau}$ changes  to $\rho_k$.
This change in the system given that the transition $i_k \to j_k$ occurred is conveniently expressed in terms of the quantum operation
\begin{equation}\label{eq:E}
\varrho_k={\mathcal E}_{j_ki_k}(\rho_{k-1})\equiv\Tr_E[\Pi_{j_k}U_k(\rho_{k-1}\otimes p_{i_k}r_{i_k})U_k^\dag\Pi_{j_k}],
\end{equation}
where $U_k$ is the time-evolution operator coupling the system to the $k$-th environment from $t=(k-1)\tau$ to  $k\tau$.
The quantum operation ${\mathcal E}$ is a completely-positive superoperator~\cite{Breuer}.
It does not preserve the trace of the system density matrix, and as such does not conserve probability.
A point we emphasize with the notation by introducing the \emph{unnormalized} density matrix $\varrho_k$ at time $t=k\tau$.
The trace of ${\mathcal E}_{j_ki_k}$ (or $\varrho_k$), instead, encodes the probability for observing the  $i_k \to j_k$ transition, $P_{j_ki_k}=\Tr_{S}[{\mathcal E}_{j_ki_k}(\rho_k)]$.
Thus, the normalized post-interaction density matrix conditioned on the transition $i_k\to j_k$ is $\rho_{k}=\varrho_k/P_{j_ki_k}$ ($={\mathcal E}_{j_ki_k}(\rho_{k-1})/P_{j_ki_k}$).
At time $t=T$, the sequence of environments is stopped, and we measure another system observable $B$ whose nondegenerate eigenvalues $b$ occur with probability $P_b=\langle b|\rho_{T^-}|b\rangle$.
The collection of measurement results we collect into a vector called the \emph{measurement trajectory} $\gamma=\{a; i_1\to j_1, \cdots, i_N \to j_N; b\}$.

To obtain the probability of observing $\gamma$ in the forward process, we first observe that the probability for each jump is the trace of ${\mathcal E}_{j_ki_k}$.
Similarly for a sequence of transitions,  we can obtain the probability of that sequence by repeated action of an ${\mathcal E}$ for each transition.
Thus, the probability to observe the sequence of transitions in $\gamma$ is
\begin{eqnarray}\label{eq:prob}
{\mathcal P}[\gamma]&=\mathcal{P}[\gamma|a]P_a \nonumber \\
\label{eq:prob2}
&=\Tr_{S}[|b\rangle\langle b|{\mathcal E}_{j_Ni_N}\cdots{\mathcal E}_{j_1i_1}|a\rangle\langle a|]P_a,
\end{eqnarray}
where it is understood that each superoperator acts on all terms to the right.

Accompanying every forward process is a \emph{reverse} process,  where each action is carried out in the reverse order and  time-reversed under the action of the time-reversal operator $\Theta$.
In particular, the initial density matrix is diagonal in the eigenbasis of ${\tilde B}=\Theta B\Theta^{-1}$, $|\tilde b\rangle=\Theta|b\rangle$, with entries $P_b$.
After first measuring ${\tilde B}$ at $t=0$, we implement the reverse protocol ${\tilde \Lambda}=\{\lambda_{\tau-t}\}_{t=0}^T$ while the system interacts with a sequence of environments, each prepared in one of the time-reversed density matrices ${\tilde r}_j=\Theta r_j\Theta^{-1}$  with probability $p_j$.
The accompanying quantum operation in the $k$-th step is
\begin{equation}
\tilde \varrho_k=\tilde{\mathcal E}_{i_kj_k}(\tilde \rho_{k-1})\equiv\Tr_E[\Pi_{i_k}{\tilde U}_k(\tilde \rho_{k-1}\otimes p_{j_k}{\tilde r}_{j_k}){\tilde U}_k^\dag\Pi_{i_k}].
\end{equation}
Here, ${\tilde U}_k$ is the reverse time-evolution operator and is related to the forward time-evolution operator by \emph{microscopic reversibility}~\cite{Campisi2010b}
\begin{equation}\label{eq:micro}
{\tilde U}_k=\Theta U_{N-k+1}^\dag\Theta^{-1},
\end{equation}
which may be verified by demonstrating that each side solves the same Schr\"odinger equation.
The processes is completed at $t=T$ by measuring ${\tilde A}=\Theta A\Theta^{-1}$.
For every trajectory of the forward process $\gamma$ there is a \emph{conjugate reverse trajectory} in which we observe the reverse sequence, $\tilde\gamma=\{{\tilde b}; j_N\to i_N, \cdots, j_1 \to i_1; {\tilde a}\}$, which occurs with probability 
\begin{eqnarray}\label{eq:probRev}
\tilde{\mathcal P}[\tilde\gamma]&=\tilde{\mathcal P}[\tilde\gamma|{\tilde b}]P_b \nonumber \\
\label{eq:probRev2}
&=\Tr_{S}[|{\tilde a}\rangle\langle {\tilde a}|\tilde{\mathcal E}_{i_1j_1}\cdots\tilde{\mathcal E}_{i_Nj_N}|{\tilde b}\rangle\langle {\tilde b}|]P_b.
\end{eqnarray}

\section{Total entropy production with repeated interactions}\label{sec:diag}

Having introduced the basic setup for modeling an open quantum systems with repeated interactions, we now investigate under what conditions we can use a detailed fluctuation theorem to identify the total entropy production.
We see from \eref{eq:dftTot} that for the detailed fluctuation theorem to hold, the ratio of ${\mathcal P}[\gamma]$ and $\tilde{\mathcal P}[\tilde\gamma]$ must depend only on $\gamma$,  that is only on the sequence of transitions between energy eigenspaces in the environment and not on the particular state to state transitions.
By inspection of \eref{eq:prob} and \eref{eq:probRev}, we have found two ways to accomplish this task.
The first, which we address in this section, is for each environment to be proportional to an energy eigenspace projection operator.
This scenario includes the case of one or many thermal reservoirs (possibly at different temperatures) and is commonly utilized in the derivations of the work fluctuation relations for unitary quantum dynamics~\cite{Campisi2010b}.
The second option is to make the effect of each environment on the system small, which we develop in \sref{sec:continuous}.

To understand how an environment proportional to an energy eigenspace projection operator leads to a detailed fluctuation theorem for the total entropy production, we expand the conditional probability to observe a trajectory $\gamma$ given the initial state $|a\rangle$ in \eref{eq:prob2} using the definition of ${\mathcal E}$  in \eref{eq:E}:
\begin{equation}
\fl
{\mathcal P}[\gamma|a]=\Tr_{SE}[|b\rangle\langle b|(\Pi_{j_N}U_N\cdots \Pi_{j_1}U_1)|a\rangle\langle a|\otimes_{k=1}^N p_{i_k}r_{i_k}(U_1^\dag\Pi_{j_1}\cdots U_N^\dag\Pi_{j_N})].
\end{equation}
When $r_i=\Pi_i/d_i$, this expression may be simplified by evaluating the trace in the energy eigenbasis of the environment.
Denoting by $\{|\alpha_{i_k}\rangle\}$ the energy eigenbasis of each eigenspace ${\mathcal H}_{i_k}$, we have
\begin{equation}\label{eq:probCond}
\fl
{\mathcal P}[\gamma|a]=\sum_{\alpha_{i_1},\alpha_{j_1},\cdots,\alpha_{i_N},\alpha_{j_N}}|\langle b,\alpha_{j_N},\dots,\alpha_{j_1}|U_N\cdots U_1|a,\alpha_{i_N},\cdots,\alpha_{i_1}\rangle|^2\frac{p_{i_N}}{d_{i_N}}\cdots\frac{p_{i_1}}{d_{i_1}}.
\end{equation}
recalling that $[U_k,\Pi_{j_l}]=0$ for $k\neq l$, since they act on different environments.
Similarly, for the reverse process we have
\begin{eqnarray}
\fl
{\mathcal P}[\tilde\gamma|{\tilde b}]&=\sum_{\alpha_{i_1},\alpha_{j_1},\cdots,\alpha_{i_N},\alpha_{j_N}}|\langle {\tilde a},\alpha_{i_1},\dots,\alpha_{i_N}|{\tilde U}_N\cdots {\tilde U}_1|{\tilde b},\alpha_{j_1},\cdots,\alpha_{j_N}\rangle|^2\frac{p_{j_1}}{d_{j_1}}\cdots\frac{p_{j_N}}{d_{j_N}} \\
\fl
&=\sum_{\alpha_{i_1},\alpha_{j_1},\cdots,\alpha_{i_N},\alpha_{j_N}}
|\langle b,\alpha_{j_N},\dots,\alpha_{j_1}|U_N\cdots U_1|a,\alpha_{i_N},\cdots,\alpha_{i_1}\rangle|^2\frac{p_{j_1}}{d_{j_1}}\cdots\frac{p_{j_N}}{d_{j_N}}.
\end{eqnarray}
where the second line follows from the antiunitarity of $\Theta$, the time-reversal invariance of $H_E$ ($\Theta |\alpha_l\rangle=|\alpha_l\rangle$), and the microscopic reversibility condition  in \eref{eq:micro}.

Now we see that the ratio of ${\mathcal P}[\gamma|a]$ and $\tilde{\mathcal P}[\tilde\gamma|{\tilde b}]$ only depends on the sequence of environment jumps, $i_k \to j_k$, as the dependence on the $\{|\alpha_{i_k}\rangle\}$ cancels.
Thus, we can identify entropy production in line with \eref{eq:dftTot} through the relation 
\begin{equation}
\Delta s_{\rm tot}[\gamma]=\Delta s[\gamma]+\Delta s_{\rm env}[\gamma]=\ln\frac{{\mathcal P}[\gamma|a]P_a}{\tilde{\mathcal P}[\tilde\gamma|{\tilde b}]P_b}
\end{equation}
with environment entropy production
\begin{equation}\label{eq:envEnt}
\Delta s_{\rm env}[\gamma]=\ln\frac{{\mathcal P}[\gamma|a]}{\tilde{\mathcal P}[\tilde\gamma|{\tilde b}]}=\sum_{k=1}^N\ln\left(\frac{p_{i_k}}{d_{i_k}}\frac{d_{j_k}}{p_{j_k}}\right).
\end{equation}
and system entropy change
\begin{equation}\label{eq:sysEnt}
\Delta s[\gamma]=-\ln P_b+\ln P_a.
\end{equation}
Notice that these results do not depend on the form of the interaction $V$ in \eref{eq:V}, and therefore apply to any interaction.

We substantiate our identifications of $\Delta s_{\rm env}$ and $\Delta s$ in \eref{eq:envEnt} and \eref{eq:sysEnt} as follows.
For an equilibrium environment, $p_j\propto d_je^{-\beta \varepsilon_j}$ and the  environment entropy production in \eref{eq:envEnt} is related to the total heat flow into the environment $q[\gamma]=\sum_k (\varepsilon_{j_k}-\varepsilon_{i_k})$ as
\begin{equation}\label{eq:entEq}
\Delta s_{\rm env}^{\rm eq}[\gamma]=\sum_{k=1}^N \beta (\varepsilon_{j_k}-\varepsilon_{i_k})=\beta q[\gamma],
\end{equation}
as expected for a thermal reservoir.
We call \eref{eq:sysEnt} the system's trajectory entropy change~\cite{Horowitz2012}, because upon averaging  over all trajectories, we find that $\Delta S=\langle \Delta s\rangle=S(\rho_T)-S(\rho_0)$ is simply the change in the von Neumann entropy $S(\rho)=-\Tr_{S}[\rho\ln\rho]$ of the system density matrix during the forward process from $\rho_0=\sum_a P_a|a\rangle \langle a|$ to $\rho_T=\sum_b P_b|b\rangle\langle b|$.
(Recall that the measurements of $A$ and $B$ force the system density matrix to be diagonal in the eigenbases of $A$ and $B$ at the beginning and end of the forward process.)

This result, however, is just a special case of the two-point measurement scheme commonly utilized in the derivation of quantum fluctuation relations with unitary dynamics for work~\cite{Campisi2010b} and currents~\cite{Esposito2009,Andrieux2009b}.
In this framework, measurements of the system and surroundings are only made at the beginning of the process, at  time $t=0$, and at the end when $t=T$.
For example, work fluctuations in open quantum systems can be identified by measuring the initial and final energies of the system and its surroundings~\cite{Campisi2010b}.
The change in system energy is the change in internal energy $\Delta U$; the change in energy of the surroundings can be equated with  the heat flow into  the system $Q$; their difference is the work $W=\Delta U-Q$.

The predictions of this section fit into this two-point measurement scheme, because \eref{eq:probCond} naturally can be interpreted as a two-point measurement.
Initially the system observable $A$ is measured along with the energy of every environment (the surroundings).
Then after a unitary evolution the observable $B$ is measured with another measurement of the energy of every environment.
Thus it makes no difference whether we measure all the environments at once upon completing the process, or measure them continually as the process evolves.

\section{Entropy production for quantum jump trajectories}\label{sec:continuous}

An alternative approach to investigating entropy production through detailed fluctuation theorems, even in the case of nonequilibrium environments, is to tune the interaction between the system and the environment so that each environment only affects the system a little.
One possibility is to make the coupling strength small, an approach often called the weak-coupling limit.
This approach we will wait to comment on till the Conclusion in \sref{sec:conclusion}.
Here, we consider another scenario typically encountered in quantum optics as a means to model continuous measurement~\cite{Kist1999,Jacobs2006,Ueda1992}, and analyzed rigorously by Nechita and Pellegrini in~\cite{Nechita2009}.
To make the environment's effect small, we take the interaction time to be short, $h\tau\ll 1$.
As a result, many environments interact with the system before it changes appreciable.
Analyzing the evolution on this longer time-scale during which the system changes, leads to an effective coarse-grained dynamics described by a quantum Markov process.
Despite the additional complications arising from passing to this limit, we find new possibilities for the types of environments we can consider and furthermore are able to develop trajectory-dependent definitions of the nonadiabatic and adiabatic entropy productions.

\subsection{Effective dynamics and the stochastic master equation}\label{sec:effective}

When the interaction time is short, it is natural to use an effective dynamics on a coarse-grained time scale that captures the effect of the rapid sequence of environments.
To achieve this limit, we take $\tau$ small with a  scaled coupling $h=\sqrt{g/\tau}$.
We then find in \ref{sec:shortApp} that on a coarse-grained time scale, with an infinitesimal increment $dt\gg \tau$ long enough that many environments interact with the system, the dynamics are well described as a Poisson process with jumps due to rare transitions in the environment.
This limit requires that we further assume  that the variation in the external parameters is not too fast, $\dot\lambda_t dt \ll 1$, and that the bare evolution generated by the system Hamiltonian $H(\lambda_t)$ over a small $dt$ interval is small.

The chance of observing a jump in the environment in any small $dt$ is very small.
When we do observe a jump, say at time $t$ from ${\mathcal H}_i\to{\mathcal H}_j$, the system density matrix $\rho_t$ changes under the action of the jump operation
\begin{equation}\label{eq:J}
\varrho_{t+dt}={\mathcal J}_{ji}(\rho_t)\equiv L_{ji}(\lambda_t)\rho_t L_{ji}^\dag(\lambda_t)dt,
\end{equation}
where the \emph{jump (or Lindblad) operators} are
\begin{equation}\label{eq:L}
L_{kl}(\lambda)=\sqrt{gp_l\langle R_{kl}^\dag R_{kl}\rangle_l}S_{kl}(\lambda),
\end{equation}
with jump (or decay) rate $g=h^2\tau$ and $\langle R_{kl}^\dag R_{kl}\rangle_l=\Tr_E[ R_{kl}^\dag R_{kl}r_l]$ is the environment correlation function.
The trace of the jump operation ${\mathcal J}_{kl}$ (or $\varrho_{t+dt}$) is the probability to observe the transition  $P_{kl}=\Tr_{S}[{\mathcal J}_{kl}(\rho)]\sim dt$, which is of order $dt$ demonstrating the anticipated Poisson statistics.
In between the jumps, say from time $s$ to $t$, the evolution is smooth and given by
\begin{equation}\label{eq:U}
\varrho_t={\mathcal U}_{t,s}(\rho_{s})\equiv U_{\rm eff}(t,s)\rho_{s} U^\dag_{\rm eff}(t,s),
\end{equation} 
where $U_{\rm eff}$ is a non-unitary \emph{effective time evolution operator} defined as the solution of (with $\hbar=1$) 
\begin{equation}\label{eq:Ueff}
\frac{\partial}{\partial t}U_{\rm eff}(t,s)=-iH_{\rm eff}(\lambda_t)U_{\rm eff}(t,s),
\end{equation}
with initial condition $U_{\rm eff}(s,s)=I$, where $I$ is the identity operator.
Here, 
\begin{equation}\label{eq:Heff}
H_{\rm eff}(\lambda)=H(\lambda)-\frac{i}{2}\sum_{kl}L_{kl}^\dag(\lambda)L_{kl}(\lambda)
\end{equation}
is a non-hermitian \emph{effective hamiltonian}.
The  operation ${\mathcal U}$ encodes the distribution of wait-times between jumps: the probability to wait from time $s$ to $t$ in state $\rho$ without observing a jump is $P_{wait}(t,s)=\Tr_{S}[{\mathcal U}_{t,s}\rho]$.

The system evolution conditioned on the environmental changes is a stochastic  process, where the smooth evolution ${\mathcal U}$ is punctuated by random jumps due to ${\mathcal J}$ that occur with Poisson statistics.
A useful alternative way to formulate this observation is in terms of a stochastic master equation for the conditioned system density matrix~\cite{Wiseman1996}.
Here the Poisson jump statistics are incorporated by introducing a collection of Poisson increments $dN_t^{kl}$, one for each type of jump between energy subspaces that can be observed.
These increments are $1$ when a jump is observed between $(t,t+dt)$ and $0$ otherwise.
Formally they are characterized by the relation $(dN^{kl}_t)^2=dN^{kl}_t$ and their classical expectation
\begin{equation}
E[dN_t^{kl}]=\Tr_{S}[L_{kl}^\dag(\lambda_t)L_{kl}(\lambda_t)\rho_t] dt.
\end{equation}
In terms of these stochastic increments the evolution captured by \eref{eq:J} and \eref{eq:U} can be given in a compact form as the It\=o stochastic differential equation for the stochastic increment of the unnormalized density matrix $d\varrho_t=\varrho_{t+dt}-\varrho_t$~\cite{Wiseman1996} 
\begin{equation}
d \varrho_t=-i(H_{\rm eff}(\lambda_t)\varrho_t - \varrho_t H_{\rm eff}^\dag(\lambda_t))dt+\sum_{kl}dN_t^{kl} (L_{kl}(\lambda_t)\varrho_t L_{kl}^\dag(\lambda_t)-\varrho_t).
\end{equation}
This equation describes the conditioned behavior of the density matrix.
The (unconditioned) average behavior of the normalized density matrix $\rho_t=\varrho_t/\Tr_{S}[\varrho_t]$ follows the Lindblad master equation~\cite{Breuer}
\begin{equation}\label{eq:master}
\frac{\partial}{\partial t} \rho_t=-i\left[H(\lambda_t),\rho_t\right]+\sum_{kl} \mathcal{D}[L_{kl}(\lambda_t)]\rho_t \equiv\mathcal{L}_{\lambda_t}\rho_t,
\end{equation}
 where $\mathcal{D}[c]\rho=c\rho c^\dag - \frac{1}{2}c^\dag c \rho - \frac{1}{2}\rho c^\dag c$.
The dynamics generated by \eref{eq:master} is a quantum Markov semigroup and is the closest analog of a classical Markov process.
It is worth noting that for a given master equation, such as \eref{eq:master}, the representation in terms of jump operators $L_{kl}$ is not unique: there exist different collections of jump operators (and $H$) that lead to the same evolution for $\rho_t$~\cite{Breuer}.
However, we have arrived at our representation using a particular microscopic model, singling out our representation with a particular measurement scheme.
Different ways of monitoring are related to other representations~\cite{Wiseman1996}.

In the reverse process, we have a similar situation.
Jumps change the system by  
\begin{equation}\label{eq:JRev}
\tilde{\mathcal J}_{ji}(\tilde\rho_t)={\tilde L}_{ji}(\tilde\lambda_t)\tilde\rho_t {\tilde L}_{ji}^\dag(\tilde\lambda_t)dt,
\end{equation}
with  \emph{reverse jump operators}
\begin{equation}\label{eq:Lrev}
{\tilde L}_{kl}(\lambda)=\Theta L_{kl}(\lambda)\Theta^{-1}
\end{equation}
derived in \ref{sec:shortApp}.
Again, between  jumps the dynamics are given by a \emph{reverse effective time-evolution operator} ${\tilde U}_{\rm eff}(t,s)$ defined as the solution of a Schr\"odinger equation, as in \eref{eq:Ueff},  except generated by the reverse effective non-hermitian Hamiltonian ${\tilde H}_{\rm eff}(\lambda)=\Theta H_{\rm eff}^\dag(\lambda)\Theta^{-1}$.
As a result, we have the time-reversal symmetry
\begin{equation}\label{eq:micro2}
{\tilde U}_{\rm eff}(t,s)=\Theta U_{\rm eff}^\dag(T-s,T-t)\Theta^{-1}.
\end{equation}

\subsection{Local detailed balance}

At this point, we nearly have all the tools to demonstrate a detailed fluctuation theorem.
The last ingredient is a time-reversal symmetry of the jump operators that can be deduced from their definitions in \eref{eq:L}
\begin{equation}\label{eq:detBal}
L_{kl}(\lambda)= L^\dag_{lk}(\lambda) e^{\Delta s_{kl}/2},
\end{equation}
where we have identified the environment entropy production per jump
\begin{equation}\label{eq:envJump}
\Delta s_{kl}=\ln\frac{p_l\langle R_{kl}^\dag R_{kl}\rangle_l}{p_k\langle R_{lk}^\dag R_{lk}\rangle_k}
=\ln\left(\frac{p_{l}}{d_{l}}\frac{d_{k}}{p_{k}}\right)+\ln\frac{\langle R_{kl}^\dag R_{kl}\rangle_ld_l}{\langle R_{lk}^\dag R_{lk}\rangle_kd_k},
\end{equation}
which differs from \eref{eq:envEnt} by the addition of the second term depending on the environment correlation functions.
For an equilibrium environment, as well as for a collection of distinct equilibrium environments, $p_k\propto d_k e^{-\beta \varepsilon_k}$ and $r_k=\Pi_k/d_k$.
As a result, the second term in \eref{eq:envJump} is zero,  and we recover the expected result $\Delta s_{kl}=\beta (\varepsilon_k-\varepsilon_l)$ [cf.~\eref{eq:entEq}].

\Eref{eq:detBal} may be viewed as an operator version of the local detailed balance condition for classical Markov jump processes relating the probability of forward and reverse jumps to the entropy production in the environment.
In light of this analogy, \eref{eq:detBal} is one of our main observations given the role that detailed balance plays in thermodynamically consistent descriptions of equilibrium and nonequilibrium processes.
\Eref{eq:detBal}, however, is valid for a wider variety of situations when $\langle R_{kl}^\dag R_{kl}\rangle_ld_l\neq \langle R_{lk}^\dag R_{lk}\rangle_kd_k$.
In this case, there is an additional \emph{thermodynamic force}
\begin{equation}\label{eq:F}
F_{kl}=\ln\frac{\langle R_{kl}^\dag R_{kl}\rangle_ld_l}{\langle R_{lk}^\dag R_{lk}\rangle_kd_k},
\end{equation}
driving the system out of equilibrium.

To clarify the origin of this force $F$, we briefly consider the simplest nontrivial example of an environment with two energy eigenspaces.
The higher energy or excited subspace ${\mathcal H}_e$ is  one dimensional and spanned by $|e\rangle$.
The second is a two-dimensional energy eigenspace $\mathcal{H}_g$ spanned by the degenerate ground states $|g\rangle$ and $|g^\prime\rangle$, with initial preparation
\begin{equation}
r_g=\left(\begin{array}{ccc}
0 & 0 & 0 \\
0 & q & s \\
0 & s^* & q^\prime
\end{array}\right),
\end{equation}
expressed in the basis $\{|e\rangle, |g\rangle,|g^\prime\rangle\}$ with $|q|^2+|q^\prime|^2=1$.
To calculate $F$, we must determine the environment correlation functions $\langle R_{kl}^\dag R_{kl}\rangle_l$.
For transitions originating in the excited subspace, we must have $\langle R^\dag_{ge}R_{ge}\rangle _ed_e=1$, since ${\mathcal H}_e$ is one dimensional.
On the other hand, for transitions from the ground eigenspace to the excited state $|e\rangle$ we are free to choose the transition operator $R_{eg}$.
The most general choice  is
\begin{equation}
R_{eg}=\left(\begin{array}{ccc}
0 & a & b \\
0 & 0 & 0 \\
0 & 0 & 0
\end{array}\right),
\end{equation}
with $a$ and $b$ arbitrary complex numbers.
A short algebraic calculation then reveals
\begin{equation}
\langle R_{eg}^\dag R_{eg}\rangle_g d_g = 2(q|a|^2+q^\prime|b|^2+a^*bs^*+ab^*s),
\end{equation}
which  upon substitution into \eref{eq:F} leads to a force 
\begin{equation}
F_{eg}=-F_{ge}=\ln [2(q|a|^2+q^\prime|b|^2+a^*bs^*+ab^*s)].
\end{equation}
The force is nonzero when there is coherence in the environment in the energy basis ($s\neq0$).
A nonzero force also arises even without coherence ($s=0$), when $q|a|^2+q^\prime|b|^2\neq 1/2$, which depends on the environment \emph{and} the interaction.
In one of our illustrative examples in \sref{sec:example}, we will further explore $F$ within the lasing-without-inversion setting where each environment has some coherence in the energy basis.

\subsection{Total entropy production}

In the description of the process given by the stochastic dynamics, the trajectory $\gamma$ can be reduced to the collection of jumps $i_k \to j_k$ and the times $t_k$ at which they occur:  $\gamma=\{a; t_1,i_1\to j_1;\dots; t_K, i_K\to j_K; b \}$.  
This is a welcome simplification, because only when there is a change in the environment is there heat flow or entropy production.
As before, the quantum operations in \eref{eq:J} and \eref{eq:U} encode the probabilities to observe such a trajectory [see \eref{eq:prob2}],
\begin{eqnarray}
{\mathcal P}[\gamma|a]&=\Tr_{S}[|b\rangle\langle b|{\mathcal U}_{T,t_K}\cdots {\mathcal J}_{j_1i_1}{\mathcal U}_{t_1,0}|a\rangle\langle a|] \nonumber \\
\label{eq:P}
&=|\langle b| U_{\rm eff}(T,t_K)\cdots L_{j_1i_1}(\lambda_{t_1})U_{\rm eff}(t_1,0)|a\rangle|^2 dt_1\cdots dt_K.
\end{eqnarray} 
For any trajectory of the forward process $\gamma=\{a; t_1,i_1\to j_1;\dots; t_K, i_K\to j_K; b \}$ we have a conjugate reverse trajectory $\tilde\gamma=\{{\tilde b}; T-t_K,j_K\to i_K;\dots; T-t_1, j_1 \to i_1; {\tilde a} \}$ that occurs in the reverse process with probability 
\begin{eqnarray}
\fl
\tilde{\mathcal P}[\tilde\gamma|{\tilde b}]&=\Tr_{S}[|{\tilde a}\rangle\langle {\tilde a}|\tilde{\mathcal U}_{T,T-t_1}\cdots \tilde{\mathcal J}_{i_Kj_K}\tilde{\mathcal U}_{T-t_K,0}|{\tilde b}\rangle\langle {\tilde b}|]\nonumber \\ 
\label{eq:Prev}
\fl
&=|\langle {\tilde a}| {\tilde U}_{\rm eff}(T,T-t_1)\cdots {\tilde L}_{i_Kj_K}(\tilde\lambda_{T-t_K}){\tilde U}_{\rm eff}(T-t_K,0)|{\tilde b}\rangle|^2 dt_1\cdots dt_K.
\end{eqnarray} 

It is now straightforward to arrive at a detailed fluctuation theorem by comparing the definitions of ${\mathcal P}$ and $\tilde{\mathcal P}$ in \eref{eq:P} and \eref{eq:Prev} after substituting in the definition of ${\tilde L}_{ji}$ in \eref{eq:Lrev} and the time-reversal symmetries in \eref{eq:micro2} and \eref{eq:detBal}:
\begin{equation}
\frac{{\mathcal P}[\gamma|a]}{\tilde{\mathcal P}[\tilde\gamma|{\tilde b}]}=e^{\Delta s_{\rm env}[\gamma]},
\end{equation}
with trajectory environment entropy change
\begin{equation}
\Delta s_{\rm env}[\gamma]=\sum_{k=1}^K \Delta s_{j_ki_k}.
\end{equation}
The total entropy production then follows as
\begin{equation}\label{eq:entTotTraj}
\Delta s_{\rm tot}[\gamma]=\Delta s[\gamma]+\Delta s_{\rm env}[\gamma]=\ln \frac{{\mathcal P}[\gamma|a]P_a}{\tilde{\mathcal P}[\tilde\gamma|{\tilde b}]P_b},
\end{equation}
which extends~\cite{Horowitz2012}.

Again, we have seen that a detailed fluctuation theorem for the total entropy production has a consistent interpretation.
When the environments are prepared with $r_l=\Pi_l/d_l$, this prediction is  a special case of the setup in \sref{sec:diag}.
The advantage comes from the possible addition of the quantum thermodynamic force $F_{kl}$ in \eref{eq:F}, and in the following section where we provide consistent definitions of the nonadiabatic and adiabatic entropy productions that would not be possible within the framework of finite-time interactions in \sref{sec:diag}.

\section{Nonadiabatic and adiabatic entropy productions}\label{sec:nonadiabatic}

The effective stochastic description of the previous section affords us the further possibility to address detailed fluctuation theorems on the level of trajectories for adiabatic and nonadiabatic entropy productions.
Like the total entropy production, $\Delta S_a$ and $\Delta S_{\rm na}$ can be determined from ratios of path probabilities.
Here, one compares the system's evolution to its dual: a time-reversed dynamics where, roughly speaking, all the dissipative stationary currents flow in reverse.
Therefore, in this section we first review some properties of the dual of a quantum Markov process in \sref{sec:dual} before identifying quantum versions of the nonadiabtic and adiabatic entropy productions in \sref{sec:dftNA}.

\subsection{Dual process}\label{sec:dual}

The dual process is defined only in the absence of driving, which we call the \emph{frozen} dynamics.
In this case, the (unconditioned) system density matrix evolves according to a master equation $\partial_t\rho_t={\mathcal L}_\lambda\rho_t$ \eref{eq:master} generated by the superoperator ${\mathcal L}_\lambda$, where $\lambda$ has a fixed value.
We further assume that this equation has a unique, positive-definite, stationary density matrix $\pi_\lambda$, given as the solution of ${\mathcal L}_\lambda\pi_\lambda=0$.

With the definition of the frozen dynamics in hand, the dual dynamics for each fixed $\lambda$ is the dynamics for which the stationary state probability of observing a reverse trajectory ${\tilde \gamma}$, ${\mathcal P}_{\rm st}^+[\tilde\gamma]$, is the same as observing the conjugate trajectory $\gamma$ in the original process: ${\mathcal P}_{\rm st}^+[\tilde\gamma]={\mathcal P}_{\rm st}[\gamma]$~\cite{Crooks2008b}.
The existence of a dual that is also a quantum Markov semigroup -- a dynamics whose averaged evolution is described by a Lindblad master equation, as in \eref{eq:master} -- is nontrivial.
However, we are assured of its existence, since we assume a finite-dimensional system and a unique, positive-definite, stationary density matrix $\pi_\lambda$~\cite{Fagnola2007}.

Like the original process the dual process is characterized by a collection of \emph{dual jump operators} $L^+_{kl}(\lambda)$ and a \emph{dual effective time-evolution operator} $U_{\rm eff}^+(t,s)$.
By comparing ${\mathcal P}_{\rm st}^+[\tilde\gamma]$ and ${\mathcal P}_{\rm st}[\gamma]$, we demonstrate in \ref{sec:dualApp} that the dual operators are 
\begin{equation}\label{eq:dualL2}
L^+_{kl}(\lambda)=\Theta \pi_\lambda^{1/2} L_{lk}^\dag(\lambda) \pi_\lambda^{-1/2}\Theta^{-1}
\end{equation}
and
\begin{equation}\label{eq:dualU2}
U^+_{\rm eff}(t,s)=\Theta \pi_\lambda^{1/2}U^\dag_{\rm eff}(T-s,T-t)\pi_\lambda^{-1/2}\Theta^{-1},
\end{equation}
which agrees with~\cite{Fagnola2007,Fagnola2008,Chetrite2012}.
We can simplify these expressions by recalling that the representation of the dynamics in terms of jump operators is not unique (see the discussion following \eref{eq:master}).
There is a special representation called the privileged representation in which the dual operators take a particularly simple form~\cite{Fagnola2007}.
This {\em privileged representation} with respect to $\pi_\lambda$ is defined by the properties
\begin{eqnarray}
\label{eq:privL}
\pi_\lambda L_{kl}(\lambda)=\varpi_{kl}(\lambda) L_{kl}(\lambda) \pi_\lambda \\
 \label{eq:commute}
[H(\lambda),\pi_\lambda]=[\sum_{kl}L_{kl}^\dag(\lambda)L_{kl}(\lambda),\pi_\lambda]=0
\end{eqnarray}
for some $\varpi_{kl}(\lambda)$, which verify $\varpi_{kl}=\varpi_{lk}^{-1}$.
These $\varpi_{kl}$ are ratios of the eigenvalues of $\pi_\lambda$:
in particular, if we expand $\pi_\lambda=\sum_n p^{\rm st}_n|e_n\rangle\langle e_n|$ in its diagonal basis, then $\varpi_{kl}(\lambda)=p^{\rm st}_n/p^{\rm st}_m$ for all states $|e_n\rangle$ and $|e_m\rangle$ connected by $L_{kl}$~\cite{Fagnola2007}.
In other words, each $L_{kl}$ only connects pairs of states whose stationary state probabilities have the same ratio.
Furthermore, the conditions in \eref{eq:commute} imply that the stationary density matrix $\pi_\lambda$ is diagonal in the system energy eigenbasis.

In the privileged representation, assuming that the jump operators are time-reversal invariant $\Theta L_{kl}\Theta^{-1}=L_{kl}$,  the dual jump operators are simply proportional to the original jump operators
\begin{equation}\label{eq:dualL}
L_{kl}^+(\lambda)=\varpi_{kl}^{1/2}(\lambda)L^\dag_{lk}(\lambda),
\end{equation}
which follows from substituting \eref{eq:privL} into \eref{eq:dualL2}.
Furthermore, we show in \ref{sec:dualApp} that \eref{eq:dualU2} for $U_{\rm eff}^+$ simplifies to 
\begin{equation}\label{eq:dualU}
U_{\rm eff}^+(t,s)=U_{\rm eff}(t,s),
\end{equation}
by verifying that both sides solve the same differential equation.
\Eref{eq:dualL} is reminiscent of the classical definition of a dual Markov process in terms of ratios of stationary state probabilities.
Furthermore, \eref{eq:dualU} is the quantum equivalent of the property that the wait-time distributions for a classical Markov process and its dual are equal.
In the following, we assume that the dynamics are in the  privileged representation.
This may seem to be a restriction; however, in the examples analyzed in \sref{sec:example} the dynamics naturally take this form, pointing to the possibility that for a quantum thermodynamic process this is generic.

\subsection{Nonadiabatic and adiabatic entropy productions}\label{sec:dftNA}

Esposito and Van den Broeck have shown that the adiabatic and nonadiabatic entropy productions can be identified as ratios of trajectory probabilities as in \eref{eq:dftNA} \cite{Esposito2010}.
We now use that connection to define the quantum analogs for quantum jump trajectories as ratios of ${\mathcal P}[\gamma]$ to the probabilities of observing $\gamma$ in the dual dynamics, ${\mathcal P}^+[\gamma]$, or $\tilde\gamma$ in the dual dynamics during the reverse process, $\tilde{\mathcal P}^+[\tilde\gamma]$:
\begin{equation}\label{eq:entTrajA}
\Delta s_{\rm na}[\gamma]=\ln\frac{{\mathcal P}[\gamma]}{{\tilde {\mathcal P}^+}[\tilde\gamma]},\qquad
\Delta s_{\rm a}[\gamma]=\ln\frac{{\mathcal P}[\gamma]}{{\mathcal P}^+[\gamma]}.
\end{equation}
Roughly speaking, $\Delta s_a$ and $\Delta s_{\rm na}$ are two measures of how distinguishable the system's evolution is from its dual.
The nonadiabatic entropy production $\Delta s_{\rm na}$ is due to the breaking of time reversal by the external driving; whereas, the adiabatic contribution $\Delta s_{\rm a}$ is due to nonequilibrium boundary conditions.

Useful formulas for $\Delta s_a$ and $\Delta s_{\rm na}$ are obtained by exploiting the symmetry in \eref{eq:dualL} and 
\begin{equation}\label{eq:db2}
{\tilde L}^+_{kl}(\lambda)=\varpi_{kl}^{1/2}(\lambda)\Theta L_{lk}^\dag(\lambda)\Theta^{-1},
\end{equation}
for the dual operators of the reverse process, which may be deduced by combining \eref{eq:Lrev} with the definition of $L^+_{kj}$ in \eref{eq:dualL}. 
Substituting this symmetry, \eref{eq:dualL}, \eref{eq:dualU} and $\varpi_{kl}=\varpi_{lk}^{-1}$ into \eref{eq:entTrajA}, we find that
\begin{eqnarray}\label{eq:sTraj}
\Delta s_{\rm na}[\gamma]&=\Delta s[\gamma]+\sum_{k=1}^K \ln\varpi_{j_ki_k}(\lambda_{t_k}) \\
\label{eq:sTraj2}
\Delta s_{\rm a}[\gamma]&=\Delta s_{\rm env}[\gamma] -\sum_{k=1}^K\ln\varpi_{j_ki_k}(\lambda_{t_k}),
\end{eqnarray}
reminiscent of their classical analogues: the $\varpi_{ji}$ have the form of ratios of stationary state probabilities, as would appear in the classical expressions.
Continuing the classical analogy, we also identify the excess entropy production~\cite{Esposito2010b}
\begin{equation}
\Delta s_{\rm ex}[\gamma]=\sum_{k=1}^K \ln\varpi_{j_ki_k}(\lambda_{t_k}).
\end{equation}

\section{Illustrations}\label{sec:example}

In this section we illustrate the nonadiabatic and adiabatic entropy productions.
Our first example is a three-level atom coupled to a coherent reservoir, inducing a nonzero quantum thermodynamic force \eref{eq:F}.
Our second example is a driven harmonic oscillator coupled to two different thermal reservoirs at inverse temperatures $\beta$ and $\beta^\prime$.

\subsection{Coherent thermal reservoir}

To demonstrate how a nonzero quantum thermodynamic force $F$ \eref{eq:F} can appear, we analyze a three-level atom coupled to a coherent thermal reservoir.
Thermodynamic consequences of adding coherence to a thermal reservoir were first discussed by Scully {\em et~al.} in the context of a photo-Carnot engine~\cite{Scully2003,Scully2002,Rostovtsev2003}.
Remarkably, the additional coherence was found to increase the overall efficiency of the engine.
Similar increases in efficiency due to coherence were observed by Dillenschneider and Lutz for a sequence of entangled environements~\cite{Lutz2009}.
Each of these studies point to the surprising effects quantum coherence in the surroundings have on the thermodynamics of a system.

\begin{figure}[htb]
\centering
\includegraphics[scale=.4]{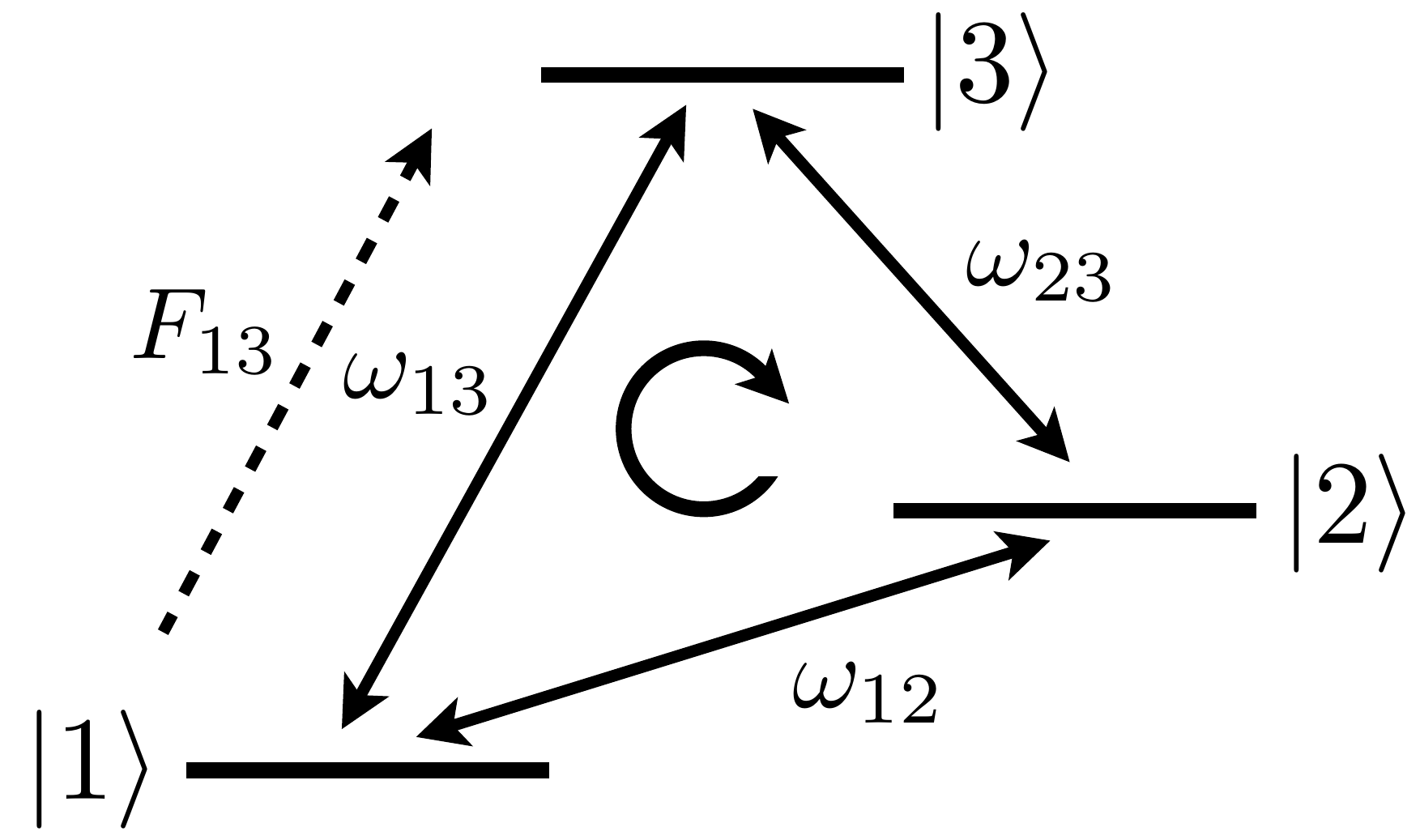}
\caption{Illustration of the three-state atom with states $|1\rangle$, $|2\rangle$, and $|3\rangle$ and transition energies $\omega_{12}$, $\omega_{23}$, and $\omega_{13}$.
A quantum thermodynamic force $F_{13}$ due to coherence in the environments coupling states $|1\rangle$ and $|3\rangle$ drives transitions (dotted arrow), generating a nonzero, clockwise current (circular arrow).}
\label{fig:3Level}
\end{figure}

\begin{figure}[htb]
\centering
\includegraphics[scale=.6]{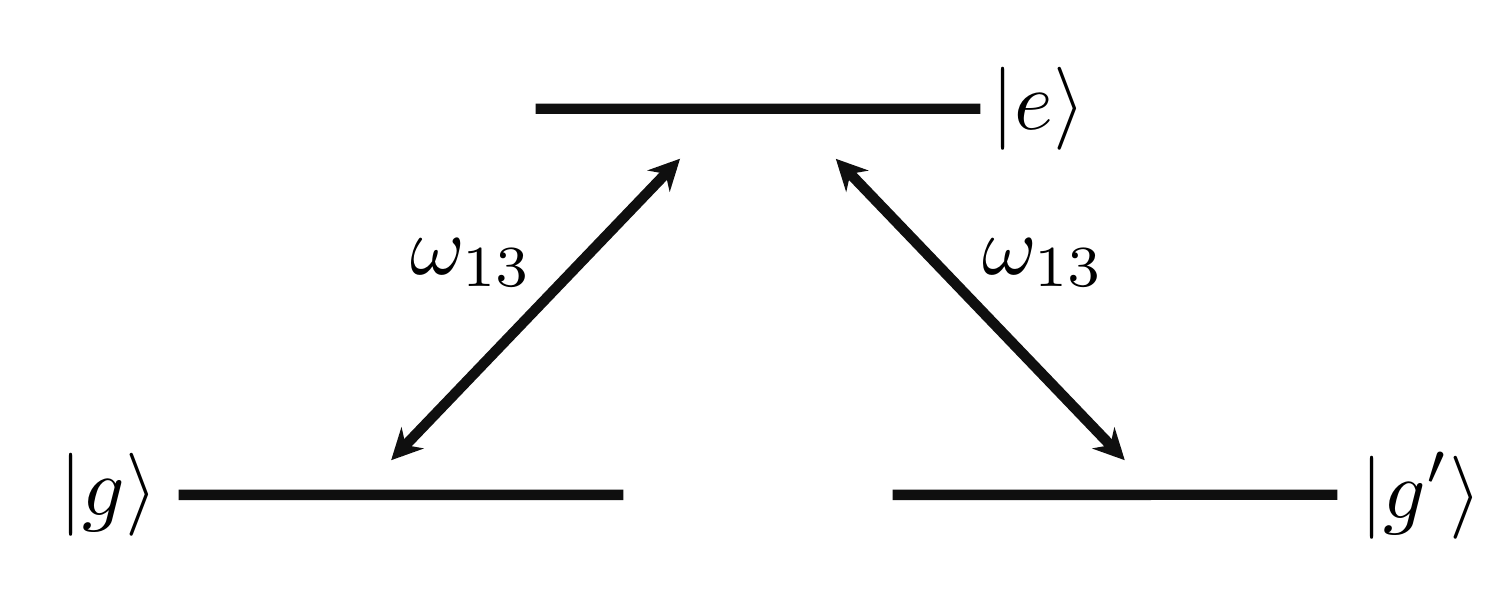}
\caption{Illustration of the three-state atoms in a $\Lambda$-configuration that mediate the $|1\rangle\leftrightarrow|3\rangle$ transition through a resonant coupling.
Coherence between the degenerate ground states $|g\rangle$ and $|g^\prime\rangle$ reduces the transition rate to the excited state $|e\rangle$ leading to the appearance of the thermodynamic force $F_{13}$.}
\label{fig:lambda}
\end{figure}

Here, we have in mind a three-level atom with states $|1\rangle$, $|2\rangle$, and $|3\rangle$ and transition energies  $\omega_{mn}>0$ ($m,n=1,2,3$), as depicted in \fref{fig:3Level}.
In this example, we will assume that the external parameters are fixed, so that we can focus on the effect of $F$.
The surroundings are implemented by coupling each transition to a rapid sequence of atoms, which play the role of our environments.
The transitions between $|1\rangle \leftrightarrow |2\rangle$ and $|2\rangle \leftrightarrow |3\rangle$ are mediated by two-level atoms resonantly coupled.
For example,  in the $|1\rangle \leftrightarrow |2\rangle$ transition,  the two-level atoms have states $|e\rangle$ and $|g\rangle$, energy difference $\omega_{12}$, and are coupled through the interaction
\begin{equation}
v_{12}=h(|2\rangle \langle 1| \sigma+|1\rangle\langle 2| \sigma^\dag),
\end{equation}
where $\sigma=|g\rangle\langle e|$.
Each atom is initially prepared in equilibrium at inverse temperature $\beta$ with $p_e/p_g=e^{-\beta\omega_{12}}$.
The $|2\rangle \leftrightarrow |3\rangle$ transition is implemented in exactly the same manner with a different stream of two-level atoms at equilibrium at inverse temperature $\beta$ with transition energy  $\omega_{23}$.
On the other hand, the $|1\rangle\leftrightarrow |3\rangle$ transitions are mediated by a stream of three-level atoms in the $\Lambda$-configuration, depicted in \fref{fig:lambda}, with excited state $|e\rangle$ and degenerate ground states $|g\rangle$ and $|g^\prime\rangle$. We set the interaction (\ref{eq:V}) for these transitions such that the states $|g\rangle$ and $|g'\rangle$ of the bath are both mapped to $|e\rangle$ when the system jumps from  $|3\rangle\rightarrow |1\rangle$, whereas
$|e\rangle$ jumps to the linear combination $\left( |g\rangle + |g'\rangle\right)/\sqrt{2}$ in a $|1\rangle\rightarrow |3\rangle$ transition.
Moreover, these three-level atoms are prepared in a coherent thermal state at inverse temperature $\beta$:
the two energy eigenspaces ${\mathcal H}_e$ and ${\mathcal H}_g$ each occur according to the Boltzmann weight $p_e/p_g=e^{-\beta \omega_{13}}/2$.
However, the density matrix in the ${\mathcal H}_g$ subspace spanned by $|g\rangle$ and $|g^\prime\rangle$ is prepared with off-diagonal elements in the energy basis according to the density matrix $r_g=\chi$.
The off-diagonal elements $\chi_{gg^\prime}=\chi_{g^\prime g}^*$ cause interference when the environment atom jumps from ${\mathcal  H}_g$ to ${\mathcal H}_e$, slowing the rate of transitions from $|3\rangle \to |1\rangle$.
This leads to a relative enhancement in the rate of $|1\rangle \to |3\rangle$ transitions accompanied by  a nonequilibrium current flowing between the different states of our three-level atom.

Carrying out the program outlined in \sref{sec:continuous}, we find that there are six jump operators $L_{kl}$ connecting the different states $|l\rangle$ to $|k\rangle$ ($k,l=1,2,3)$.
Here,  we label the jump operators by state transitions instead of environment transitions, because each environment transition naturally corresponds to only one state transition.
The six jump operators  satisfy local detailed balance \eref{eq:detBal}, 
\begin{eqnarray}
L_{13}=e^{(\beta\omega_{13}+F_{13})/2}L_{31}^\dag \\
L_{23}=e^{\beta\omega_{23}/2}L_{32}^\dag \\
L_{12}=e^{\beta\omega_{12}/2}L_{21}^\dag,
\end{eqnarray}
where 
\begin{equation}
F_{13}=\ln(1+\chi_{gg^\prime}+\chi_{g^\prime g})=\ln(1+2{\rm Re}\chi_{gg^\prime}),
\end{equation}
Clearly, the force is only nonzero when there is some coherence in the energy eigenbasis (${\rm Re} \chi_{gg^\prime}\neq 0$).

The nonzero value of $F$ leads to entropy production as follows.
Since the dynamics are autonomous -- the external parameters are fixed -- the system eventually relaxes to a steady state that is diagonal in the energy basis.
At which point, the force $F_{13}$ continually drives transitions $|1\rangle\to|3\rangle\to|2\rangle$ resulting in a nonzero stationary current $J_{\rm st}$, running clockwise in \fref{fig:3Level}.
In this case, the average total entropy production rate takes the familiar form of force times flux, ${\dot S}_{\rm tot}=F_{13}J_{\rm st}$.

In their original analysis of the photo-Carnot engine, Scully {\em et al.} did not identify the thermodynamic force $F$~\cite{Scully2003}.
Instead they assigned a modified temperature $T_{\rm mod}$ to the coherent thermal reservoir that depended on the degree of coherence.
This modified temperature was identifiable because they studied the photo-Carnot engine where the working substance is a single mode of the electromagnetic field, and when coupled to the coherent reservoir it relaxes to a Boltzmann distribution at the modified temperature $T_{\rm mod}$.
However, in our three-state atom model no such conclusion exists.  
The coherence maintains  the system away from equilibrium leading to a constant entropy production.

\subsection{Driven harmonic oscillator}

In our second example, we present  $\Delta s_{\rm na}$ and $\Delta s_{\rm a}$ for a driven quantum harmonic oscillator coupled to thermal reservoirs at different temperatures.
The oscillator's Hamiltonian in terms of its position $x$, momentum $p$, frequency $\omega$, and mass $m$ in units with $\hbar=1$ is
\begin{equation}\label{eq:harHam}
H(\lambda)=\frac{p^2}{2m}+\frac{1}{2}m\omega(x-\lambda)^2
\end{equation}
and is parameterized by the external parameter $\lambda$, which translates the minimum of the harmonic well.
This motivates introducing the translated raising and lowering operators
\begin{eqnarray}
a_\lambda=\sqrt{\frac{m\omega}{2}}\left(x-\lambda+\frac{ip}{m\omega}\right) \\
a^\dag_\lambda=\sqrt{\frac{m\omega}{2}}\left(x-\lambda-\frac{ip}{m\omega}\right),
\end{eqnarray}
which when substituted into the Hamiltonian \eref{eq:harHam} lead to the simplified expression
\begin{equation}
H(\lambda)=\omega(a^\dag_\lambda a_\lambda+1/2).
\end{equation}
To do work on the oscillator, we vary $\lambda$ according to a protocol $\Lambda=\{\lambda_t \}_{t=0}^T$ from $t=0$ to $T$.
In order to use the effective dynamics outlined in \sref{sec:continuous}, we assume that the driving is not too fast, $\dot\lambda_t dt\ll1$.
For notational simplicity, we will denote the translated raising and lowering operators at time $t$ when the external parameter has value $\lambda_t$ as $a_t\equiv a_{\lambda_t}$ and $a^\dag_t\equiv a^\dag_{\lambda_t}$

The surroundings are composed of two distinct sequences of two-level atoms, which will mimic two thermal reservoirs at inverse temperatures $\beta$ and $\beta^\prime$.
Each atom of the  $\beta$ ($\beta^\prime$) reservoir has  ground state  $|g\rangle$ ($|g^\prime\rangle$) and excited state $|e\rangle$ ($|e^\prime\rangle$), prepared with the Boltzmann weight $p_e/p_g=e^{-\beta\omega}$ ($p_e^\prime/p_g^\prime=e^{-\beta^\prime\omega}$).
The interaction is
\begin{equation}
V(\lambda)=h(a_\lambda \sigma^\dag +a_\lambda^\dag\sigma)
\end{equation}
where $\sigma=|g\rangle\langle e|$ for the $\beta$ reservoir or  $\sigma=|g^\prime\rangle\langle e^\prime|$ for the atoms at $\beta^\prime$.

With this setup, there are four jump operators,
two for the $\beta$ reservoir,
\begin{equation}\label{eq:harJump}
L_{eg}(\lambda_t)=\sqrt{gp_g}a_t\qquad L_{ge}(\lambda_t)=\sqrt{g p_e} a^\dag_t,
\end{equation}
and two for the $\beta^\prime$ reservoir,
\begin{equation}\label{eq:harJump2}
L^\prime_{eg}(\lambda_t)=\sqrt{g p^\prime_g}a_t\qquad L^\prime_{ge}(\lambda_t)=\sqrt{g p^\prime_e} a^\dag_t.
\end{equation}
Each pair satisfies  detailed balance \eref{eq:detBal}:
\begin{equation}\label{eq:detBalhar}
L_{eg}(\lambda)=L^\dag_{ge}(\lambda)e^{\beta\omega/2}\qquad L^\prime_{eg}(\lambda)=[L^\prime_{ge}(\lambda)]^\dag e^{\beta^\prime\omega/2},
\end{equation}
where we can identify the entropy production per jump in the $\beta$ reservoir $\Delta s_{eg}=\beta\omega$ and in the $\beta^\prime$ reservoir $\Delta s_{eg}^\prime=\beta^\prime\omega$, as  the heat transferred to the reservoir divided by the temperature, as expected for a thermal reservoir.

Furthermore, these dynamics support a unique stationary state for each fixed $\lambda$, 
\begin{equation}\label{eq:betaEff}
\pi_\lambda=\frac{1}{Z_\lambda}e^{-\beta_{\rm eff}H(\lambda)}
\end{equation}
where $Z_\lambda=\Tr\exp[-\beta_{\rm eff}H(\lambda)]$ and 
\begin{equation}
\beta_{\rm eff}=-(1/\omega)\ln\left[(p_e+p_e^\prime)/(p_g+p_g^\prime)\right].
\end{equation}
Even though $\pi_\lambda$ \eref{eq:betaEff} resembles an equilibrium density matrix, the oscillator is not in equilibrium as heat flows through the system from the hotter to cooler reservoir.
The associated dual jump operators, which are obtained by substituting \eref{eq:harJump}, \eref{eq:harJump2} and \eref{eq:betaEff} into \eref{eq:dualL2}, are
\begin{equation}\label{eq:dualHar}
L^+_{eg}(\lambda)=e^{\beta_{\rm eff}\omega/2}L^\dag_{ge}(\lambda)\qquad L^+_{ge}(\lambda)=e^{-\beta_{\rm eff}\omega/2}L^\dag_{eg}(\lambda)
\end{equation}
and 
\begin{equation}
[L^\prime_{ge}(\lambda)]^+=e^{\beta_{\rm eff}\omega/2}[L_{eg}^\prime(\lambda)]^\dag \qquad [L^\prime_{eg}(\lambda)]^+=e^{-\beta_{\rm eff}\omega/2}[L_{ge}^\prime(\lambda)]^\dag.
\end{equation}
Notice that despite the harmonic oscillator being infinite dimensional, the dual operators are proportional to the original jump operators, allowing us to still consider the adiabatic and nonadiabtic entropy productions.

With this setup it is now straightforward to determine expressions for the nonadiabatic and adiabatic entropy productions.
The forms are most suggestive in terms of the heat flow into the $\beta$ ($\beta^\prime$) thermal reservoir along the trajectory $\gamma$, $q[\gamma]$ ($q^\prime[\gamma]$).
Combing \eref{eq:detBalhar}, \eref{eq:dualHar}, \eref{eq:sTraj} and \eref{eq:sTraj2}, we have
\begin{eqnarray}
\Delta s_a[\gamma]=(\beta-\beta_{\rm eff})q[\gamma]+(\beta^\prime -\beta_{\rm eff})q^\prime[\gamma] \\ 
\Delta s_{\rm na}[\gamma]=\Delta s[\gamma]+\beta_{\rm eff}(q[\gamma]+q^\prime[\gamma]).
\end{eqnarray}

For a linear driving with rate $\nu$, $\lambda_t=\nu t$, the system relaxes to a steady state in the long time limit, because in the co-moving frame the system appears at rest. 
In this steady state, entropy is continually being produced.
It is then more informative to analyze the rate of entropy production.
In this case, the system entropy change $\Delta s[\gamma]$ vanishes on average, energy conservation guarantees that the average rates of heat flow balance, ${\dot Q}=-{\dot Q}^\prime$, and the average entropy production rates become ${\dot S}_{\rm na}=0$ and ${\dot S}_{\rm tot}= {\dot S}_a=\Delta \beta {\dot Q}$, where $\Delta \beta=\beta-\beta^\prime$.

\section{Conclusion}\label{sec:conclusion}

For a wide class of open quantum systems evolving along quantum jump trajectories, we have established formulas for the trajecotry-dependent total \eref{eq:entTotTraj}, adiabatic \eref{eq:sTraj}, and nonadiabatic \eref{eq:sTraj2} entropy productions by means of detailed fluctuation theorems.
To arrive at these formulas, we used the continuous measurement framework in which each environment interacts with the system for a short time $\tau$.
To achieve this limit, we required that the coupling strength scale as $h\sim 1/\sqrt{\tau}$.
Thus, as  $\tau\to 0$ the coupling becomes infinite, yet the effect on the system remains small.
A more physical limit where the coupling is finite is the weak coupling limit~\cite{Breuer}, where $h$ is small and the interaction time is long.
Unfortunately, the weak-coupling limit applies to a more restrictive class of protocols.
To these authors knowledge only periodic~\cite{Breuer1997}, linear~\cite{Horowitz2012} and slow ~\cite{Alicki1979} driving have been successfully incorporated.
Therefore, despite the unphysical nature of the continuous measurement framework, we can consider a larger variety of nonequilibrium drivings.

The key to our analysis is the local detailed balance relation \eref{eq:detBal}, connecting entropy production to a symmetry of the jump operators.
This relation is a natural consequence of the time-reversal symmetry of the underlying microscopic dynamics.
Often in classical statistical thermodynamics local detailed balance is considered a minimal requirement for a consistent thermodynamic description~\cite{Seifert2012}.
Similarly, one could take \eref{eq:detBal} as a starting point for a consistent description of the thermodynamics of jump trajectories.
Beyond the classical regime, this relation holds even when the surroundings are coherent in the energy eigenbasis, leading to the appearance of a quantum thermodynamic force $F$ \eref{eq:F}.
Unfortunately, there is no clear connection between this force and the change in Shannon entropy of the surroundings as one would hope.
In general, it remains an open problem to interpret the environment entropy production \eref{eq:envJump} for any nonthermal environment.
We suspect that if we embed such a nonthermal environment in a larger thermal super-reservoir, then \eref{eq:envJump} may bound the physical entropy production in the super-reservoir.
This intuition is supported by the analysis of Scully~\cite{Scully2002} and Zubairy~\cite{Zubairy2002} showing that the work necessary to create the coherent reservoir exploited by the photo-Carnot engine always exceeds the useful work extracted from the engine.
Despite this shortcoming, the introduction of $F$ still allows for a consistent description of irreversibility -- the distinguishability of the forward and reverse processes -- through the total entropy production.

Our analysis has focused on entropy production and its fluctuations.
However, the setup we have presented here can be extended to incorporate a variety fluctuation relations.
The key observation  is that instead of initializing the reverse process with the final distribution of the forward process, $P_b$, we can use any other probability distribution ${\tilde P}_b$.
Then the trajectory functional ${\mathcal R}[\gamma]=\ln(\mathcal{P}[\gamma|a]/\tilde{\mathcal P}[\tilde\gamma|{\tilde b}])+\ln P_a/{\tilde P_b}$ will satisfy an integral and detailed fluctuation theorem.
By making various choices for the operator $B$ and ${\tilde P}_b$, we can arrive at any of a collection of fluctuation relations as in the classical case~\cite{Harris2007,Seifert2005b}, see for example the discussion in \cite{Horowitz2012} on work fluctuation relations.

In contrast to the trajectory analysis we have performed, Chetrite and Mallick have recently predicted an integral fluctuation theorem for the nonadiabatic entropy production for driven quantum Markov semigroups~\cite{Chetrite2012}.
Their result is based on quantum multi-time correlation functions and makes no explicit mention of trajectories.
While we expect that their results and ours are closely related, the precise relationship is still an open question that deserves further analysis.

To conclude, we must comment on the classical nature of the quantum jump trajectories we have discussed.
As merely a collection of classical measurement outcomes, they are classical trajectories.
However, each trajectory can be mapped to a stochastic path through Hilbert space traced out by the conditioned wave vector of the system \cite{Horowitz2012}.
Such an identification is possible, since the interaction in \eref{eq:V} has only one term that mediates each transition between environment energy eigenspaces.
As a result, the jump operations in \eref{eq:J} map pure states to pure states, allowing an evolution described completely in terms of vectors.
For a more general interaction, such an identification would not be possible.
Instead, one may consider the evolution of the conditioned density matrix as a quantum trajectory.

Despite the classical nature of the measurement trajectories, quantum effects still manifest themselves in their statistics.
In particular, such quantum influence is observed when there is coherence in the surroundings, or when the driving is rapid enough to cause nonadiabatic transitions between between energy eigenstates of the system.
Even more opportunities exist to use the framework outlined here to study the thermodynamics of a variety of purely quantum phenomena such as decoherence, entanglement, or measurement.

\ack
JMH is supported financially by the National Science Foundation (USA) International Research Fellowship under Grant No. OISE-1059438 and by ARO MURI grant W911NF-11-1-0268.
JMRP is supported by the Spanish Government under FIS2011-22644.

\appendix

\section{Effective dynamics for short interactions}\label{sec:shortApp}

In this appendix, we demonstrate how to arrive at  the effective dynamics outlined in \sref{sec:effective} when the interaction time is short $h\tau\ll1$ and the coupling strength  scales as $h=\sqrt{g/\tau}$.
The technique we apply here is adapted from \cite{Kist1999,Ueda1992}.

To begin, we  solve the Schr\"odinger equation for the time-evolution operator $U(t,t_0)$ during one $\tau$-interval starting at $t=t_0$ for small times $h\tau\ll 1$:
\begin{equation}
\frac{\partial U}{\partial t}=-i(H(\lambda_t)+H_E+V(\lambda_t))U,
\end{equation}
with initial condition $U(t_0,t_0)=I$.
This is facilitated by entering the interaction picture by multipling $U$ by the operator 
\begin{equation}
K(t,t_0)=K_{S}(t,t_0)K_E(t,t_0)\equiv{\mathcal T}\exp\left[-i\int_{t_0}^t ds\, H(\lambda_{s})\right]\exp[-i (t-t_0)H_E]
\end{equation}
to get $U_I(t,t_0)=K^\dag(t,t_0)U(t,t_0)$, which evolves according to
\begin{equation}\label{eq:SEinter}
\frac{\partial U_I}{\partial t}=-i V_I(t) U_I,
\end{equation}
where $V_I(t)=K^\dag (t,t_0)V(\lambda_t)K(t,t_0)$.
A formal solution to \eref{eq:SEinter} is given by the series
\begin{equation}
U_I(t_0+\tau,t_0)=I+\sum_{n=1}^\infty (-i)^n\int_{t_0}^\tau ds_n \cdots \int_{t_0}^{s_2} ds_1\, V_I(s_n)\cdots V_I(s_1),
\end{equation}
whose approximate solution for small times $\tau$ is
\begin{equation}\label{eq:Uapprox}
U_I(t_0+\tau,t_0)\approx I-i\tau V_I(t_0)-\frac{\tau^2}{2}(V_I^2(t_0)+i \partial_t V_I(t_0)).
\end{equation}

\Eref{eq:Uapprox} now allows us to deduce the short time behavior for the quantum operation ${\mathcal E}_{ji}$ when the environment transitions from ${\mathcal H}_i$ to a \emph{different} eigenspace ${\mathcal H}_j$.
Exploiting the property that each $R$ only connects one pair of energy eigenspaces of the environment, and $[K_E,H_E]=0$, gives
\begin{equation}
{\mathcal E}_{ji}(\rho)\approx \tau^2 K_S(t_0+\tau,t_0)\Tr_E[\Pi_jV_I(t_0)\rho\otimes p_i r_i V_I^\dag(t_0)]K_S^\dag(t_0+\tau,t_0).
\end{equation}
Substituting in the definitions of $V$ in \eref{eq:V} and $L_{ji}$ in \eref{eq:L}, and $h=\sqrt{g/\tau}$ leads to
\begin{equation}
{\mathcal E}_{ji}(\rho)\approx \tau K_{S}(t_0+\tau,t_0)L_{ji}(\lambda_{t_0})\rho L^\dag_{ji}(\lambda_{t_0})K_{S}^\dag(t_0+\tau,t_0).
\end{equation}
Similarly, when the environment does not transition
\begin{eqnarray}
\fl
{\mathcal E}_{ii}(\rho)&\approx &K_{S}(t_0+\tau,t_0)p_i \rho K^\dag_{S}(t_0+\tau,t_0) \\
\fl
&\quad&-\frac{\tau^2}{2}K_S(t_0+\tau,t_0)\Tr_E\left[\Pi_i\left(V_I^2(t_0)\rho\otimes p_i r_i+\rho\otimes p_ir_i (V_I^\dag(t_0))^2\right)\right]K_S^\dag(t_0+\tau,t_0)\\
\fl
&\approx& K_{S}(t_0+\tau,t_0)\left[p_i \rho-\frac{\tau}{2}\{\sum_l L^\dag_{li}(\lambda_{t_0})L_{li}(\lambda_{t_0}),\rho\}\right]K_{S}^\dag(t_0+\tau,t_0),
\end{eqnarray} 
where for two operators $O$ and $O^\prime$, $\{O,O^\prime\}=OO^\prime+O^\prime O$ is the anticommutator.

We now investigate the evolution during a small time interval $dt=M\tau$ with $M\gg 1$, during which many environments interact with the system, but the probability to observe a jump remains small.
This coarse-graining allows us to describe the dynamics as a Poisson process, where the jumps are rare and occur instantaneously on the time scale $\tau \ll dt$.
Furthermore, we assume that during $dt$ the system dynamics are negligible, and we can approximate $K_S(t_0+dt,t_0)\approx I-i H(\lambda_{t_0}) dt$.

Consider a $dt=M\tau$ window from $t=t_0$ to $t_0+dt$ when a jump is observed, say $i \to j$.
We cannot know precisely which of the $M$ environments jumped.
Therefore, to correctly predict the change in the system, we must sum over all possibilities
\begin{eqnarray}
{\mathcal J}_{ji}(\rho)&=\sum_{k=1}^M \sum_{i_M}{\mathcal E}_{i_Mi_M}\cdots{\mathcal E}_{j_ki_k}\cdots \sum_{i_1}{\mathcal E}_{i_1i_1}(\rho) \\
\label{eq:jumpDT}
&\approx L_{ji}(\lambda_{t_0})\rho L_{ji}^\dag(\lambda_{t_0})dt,
\end{eqnarray} 
keeping only terms lowest order in $dt$, and remembering that the external parameters $\lambda$ do not change significantly during $dt$.
The probability for this event is 
\begin{equation}\label{eq:probJump}
P_{ji}=\Tr_{S}[{\mathcal J}_{ji}(\rho)]=\langle L_{ji}^\dag(\lambda_{t_0})L_{ji}(\lambda_{t_0})\rangle dt,
\end{equation}
where $\langle \cdot\rangle=\Tr_{S}[\cdot\rho]$.
\Eref{eq:probJump} is of order $dt$ signaling that the jump statistics are described by a Poisson process.
Similarly, during an interval when no jumps are detected in any of the $M$ environments the effect on the system is given by
\begin{eqnarray}
\fl
{\mathcal U}_{t_0+dt,t_0}(\rho)&=\sum_{i_M}{\mathcal E}_{i_Mi_M}\cdots \sum_{i_1}{\mathcal E}_{i_1i_1}(\rho) \\
\fl
&\approx \left[K_{S}(t_0+dt,t_0) - \frac{dt}{2}\sum_{ji}L^\dag_{ji}(\lambda_{t_0})L_{ji}(\lambda_{t_0})\right]\rho\left[K_{S}^\dag(t_0+dt,t_0) - \frac{dt}{2}\sum_{ji}L^\dag_{ji}(\lambda_{t_0})L_{ji}(\lambda_{t_0})\right] \\ 
\fl
&\approx[I-idt H_{\rm eff}(\lambda_{t_0})]\rho[I+idt H^\dag_{\rm eff}(\lambda_{t_0})] \\
\fl
\label{eq:UeffDT}
& \approx U_{\rm eff}(t_0+dt,t_0)\rho U^\dag_{\rm eff}(t_0+dt,t_0),
\end{eqnarray}
where we have expanded $K_S(t_0+dt,t_0)\approx I-i H(\lambda_{t_0})dt$ and substituted in the definitions of $H_{\rm eff}$ \eref{eq:Heff} and $U_{\rm eff}$ \eref{eq:Ueff}.
The probability of not observing any jumps is 
\begin{equation}
P_{no-jump}=\Tr_{S}[{\mathcal U}_{t_0+dt,t_0}(\rho)]\approx 1 -\sum_{ji} \langle L_{ji}^\dag(\lambda_{t_0})L_{ji}(\lambda_{t_0})\rangle dt,
\end{equation}
which confirms that the dynamics are governed by a compound Poisson process for the different possible transitions between energy eigenspaces.
Equations~(\ref{eq:jumpDT}) and (\ref{eq:UeffDT}) demonstrate that the effective equations of motion are governed by $U_{\rm eff}$ between jumps and the operators $L_{ji}$ at jumps.

In the reverse process, the same arguments lead to a jump quantum operation at time $t$:
\begin{equation}\label{eq:JrevApp}
\tilde{\mathcal J}_{ji}(\rho)=gp_i \Tr_E[R_{ji}^\dag R_{ji}\tilde{r}_i]S_{ji}(\tilde\lambda_t)\rho S_{ji}^\dag(\tilde\lambda_t)dt\equiv {\tilde L}_{ji}(\tilde\lambda_t)\rho {\tilde L}_{ji}^\dag(\tilde\lambda_t)dt.
\end{equation}
To arrive at \eref{eq:Lrev} for $\tilde{L}_{kl}$, we first separate the time-reversal operator $\Theta=\theta_{S}\otimes \theta_E$ into the time-reversal operator on the system $\theta_{S}$ and environment $\theta_E$.
Then,
\begin{equation}
\tilde{\mathcal J}_{ji}(\rho)=gp_i \Tr_E[ R_{ji}S_{ji}(\tilde\lambda_t)\rho \otimes \theta_E{r}_i\theta_E^{-1}S_{ji}^\dag(\tilde\lambda_t)R_{ji}^\dag]dt.
\end{equation}
and its time-reversal is
\begin{equation}
\theta_{S}\tilde{\mathcal J}_{ji}(\theta_{S} \rho\theta^{-1}_{S})\theta^{-1}_{S}=gp_i \Tr_E[\Theta R_{ji}S_{ji}(\tilde\lambda_t)\Theta \rho \otimes  r_i \Theta^{-1} S_{ji}^\dag(\tilde\lambda_t)R_{ji}^\dag \Theta^{-1}],
\end{equation}
after inserting a $\theta_E\theta_E^{-1}=I$ outside the the time-reversal invariant environment operators $R_{ji}$.
Noting the time-reversal invariance of the interaction $\Theta V\Theta=V$, we arrive at
\begin{equation}
\theta_{S}\tilde{\mathcal J}_{ji}(\theta_{S} \rho\theta^{-1}_{S})\theta^{-1}_{S}=gp_i \Tr_E[R_{ji}^\dag R_{ji}{r}_i]S_{ji}(\tilde\lambda_t)\rho S_{ji}^\dag(\tilde\lambda_t),
\end{equation}
from which it is straightforward to arrive at \eref{eq:Lrev} by comparing with \eref{eq:JrevApp}.

\section{Operators for dual quantum markov processes}\label{sec:dualApp}

The dual process is defined such that ${\mathcal P}^+_{\rm st}[\tilde\gamma]={\mathcal P}_{\rm st}[\gamma]$.
To identify the jump operators $L^+_{kl}$ and effective time-evolution operator $U^+_{\rm eff}$ of the dual process, we begin by analyzing ${\mathcal P}_{\rm st}[\gamma]$.
Since $\lambda$ is fixed throughout the calculation we will suppress it in the notation. 
For our purposes here it is sufficient to consider a trajectory with only one jump $\gamma=\{e_k;t_1,i\to j; e_l\}$ between two eigenstates $|e_k\rangle$ and $|e_l\rangle$ of the stationary density $\pi=\sum_n p^{\rm st}_n | e_n\rangle\langle e_n|$:
\begin{equation}\label{eq:ps}
{\mathcal P}_{\rm st}[\gamma]=|\langle e_l|U_{\rm eff}(T,t_1)L_{ji}U_{\rm eff}(t_1,0)|e_k\rangle|^2 p^{\rm st}_k
\end{equation}
First, we note that since $\pi$ is positive-definite it has an inverse  $\pi^{-1}$ and square root $\pi^{1/2}$.
Therefore, we  can rewrite \eref{eq:ps} as
\begin{equation}
{\mathcal P}^+_{\rm st}[\gamma]=|\langle e_l|\pi^{-1/2}U_{\rm eff}(T,t_1)L_{ji}U_{\rm eff}(t_1,0)\pi^{1/2}|e_k\rangle|^2p_l^{\rm st}.
\end{equation}
Next, following Crooks~\cite{Crooks2008b}, we insert between each pair of operators the identity operator $I=\pi^{1/2}\pi^{-1/2}$,
\begin{equation}
\fl
{\mathcal P}_{\rm st}[\gamma]=|\langle e_l|\pi^{-1/2} U_{\rm eff}(T,t_1)\pi^{1/2}\pi^{-1/2}L_{ji}\pi^{1/2}\pi^{-1/2}U_{\rm eff}(t_1,0)\pi^{1/2}|e_k\rangle|^2 p^{\rm st}_l,
\end{equation}
followed by $I=\Theta \Theta^{-1}$,
\begin{equation}
\fl
{\mathcal P}_{\rm st}[\gamma]=|\langle e_l|\Theta(\Theta^{-1}\pi^{-1/2} U_{\rm eff}(T,t_1)\pi^{1/2}\Theta)(\Theta^{-1}\pi^{-1/2}L_{ji}\pi^{1/2}\Theta)(\Theta^{-1}\pi^{-1/2}U_{\rm eff}(t_1,0)\pi^{1/2}\Theta)\Theta^{-1}|e_k\rangle|^2 p^{\rm st}_l.
\end{equation}
Finally, we use the anti-unitarity of $\Theta$ to flip the order of the operators
\begin{equation}
\fl
{\mathcal P}_{\rm st}[\gamma]=|\langle \tilde{e}_k|(\Theta\pi^{1/2} U^\dag_{\rm eff}(t_1,0)\pi^{-1/2}\Theta^{-1})(\Theta\pi^{1/2}L^\dag_{ji}\pi^{-1/2}\Theta^{-1})(\Theta\pi^{1/2}U^\dag_{\rm eff}(T,t_1)\pi^{-1/2}\Theta^{-1})|{\tilde e}_l\rangle|^2 p^{\rm st}_l,
\end{equation}
recalling that $|\tilde{e}_n\rangle=|\Theta e_n\rangle$.
Comparing with the dual process
\begin{equation}
{\mathcal P}_{\rm st}^+[\tilde\gamma]=|\langle {\tilde e}_k|U^+_{\rm eff}(T,T-t_1)L^+_{ij}U^+_{\rm eff}(T-t_1,0)|{\tilde e}_l\rangle|^2 p^{\rm st}_l,
\end{equation}
we arrive at \eref{eq:dualL2} and \eref{eq:dualU2}, reprinted here for convince 
\begin{equation}\label{eq:dualLApp}
L^+_{ij}=\Theta \pi^{1/2} L_{ji}^\dag \pi^{-1/2}\Theta^{-1}
\end{equation}
and
\begin{equation}\label{eq:dualUApp}
U^+_{\rm eff}(t,s)=\Theta \pi^{1/2}U^\dag_{\rm eff}(T-s,T-t)\pi^{-1/2}\Theta^{-1}.
\end{equation}

\Eref{eq:dualU} for $U^+_{\rm eff}$ can be derived by first differentiating with time \eref{eq:dualUApp}
\begin{eqnarray}
\frac{\partial}{\partial t} U^+_{\rm eff}(t,s)&=\Theta \pi^{1/2}\frac{\partial}{\partial t}[U^\dag_{\rm eff}(T-t,t-s)]^{-1} \pi^{-1/2}\Theta^{-1} \\
&=\Theta \pi^{1/2} (iH^\dag_{\rm eff})[U^\dag_{\rm eff}(T-t,t-s)]^{-1}\pi^{-1/2}\Theta^{-1} \\
&=-iH_{\rm eff}^+U_{\rm eff}^+(t,s)
\end{eqnarray}
where
\begin{equation}
H^+_{\rm eff}=\Theta \pi^{1/2}H^\dag_{\rm eff}\pi^{-1/2}\Theta^{-1}=\Theta \pi^{1/2}\left[H+\frac{i}{2}\sum_{kl}L_{kl}^\dag L_{kl}\right]\pi^{-1/2}\Theta^{-1}
\end{equation}
Finally, in the privileged representation $[H,\pi]=[\sum_{kl}L^\dag_{kl}L_{kl},\pi]=0$ \eref{eq:commute}, $H$ and $L_{kl}$ are assumed time-reversal invariant, and $\Theta i=-i \Theta$ implying that $H^+_{\rm eff}=H_{\rm eff}$.
Consequently, $U^+_{\rm eff}$ and $U_{\rm eff}$ satisfy the same differential equation \eref{eq:Ueff} with the same initial conditions and  therefore must be equal.

\bibliographystyle{iopart-num}
\bibliography{FluctuationTheory,PhysicsTexts,QuantumWork}

\end{document}